\documentclass[letterpaper]{article}

\usepackage{natbib}  
\usepackage{graphicx}

\usepackage{soul}
\usepackage{url}
\usepackage[hidelinks]{hyperref}
\usepackage[utf8]{inputenc}
\usepackage[small]{caption}
\usepackage{graphicx}
\usepackage{amsmath}
\usepackage{booktabs}
\usepackage{algorithm}
\usepackage{algorithmic}
\usepackage{amsmath}
\usepackage{color} 
\usepackage{float}
\usepackage{hyperref}

\usepackage{setspace} 

\usepackage{amssymb}
\usepackage{graphicx}
\usepackage[scale=0.75, top=2cm]{geometry}
\usepackage{enumerate}
\usepackage{siunitx,booktabs}

\hypersetup{
    bookmarks=true,         
    unicode=false,          
    pdftoolbar=true,        
    pdfmenubar=true,        
    pdffitwindow=false,     
    pdfstartview={FitH},    
    pdftitle={My title},    
    pdfauthor={Author},     
    pdfsubject={Subject},   
    pdfcreator={Creator},   
    pdfproducer={Producer}, 
    pdfkeywords={keywords}, 
    pdfnewwindow=true,      
    colorlinks=true,       
    linkcolor=blue,          
    citecolor=blue,        
    filecolor=magenta,      
    urlcolor=cyan           
    }

\usepackage{todonotes}
\usepackage{caption}
\usepackage{booktabs}
\usepackage{float}

\date{}

\title{Evolution of Coordination in Pairwise and Multi-player Interactions via Prior Commitments }
\author{Ndidi Bianca Ogbo$^{1}$, Aiman Elgarig$^{1}$,  The Anh Han$^{1,\star}$ \\
\mbox{}\\
$^1$School of Computing, Engineering and Digital Technologies, Teesside  University \\
$^\star$ Corresponding author: The Anh Han (han@tees.ac.uk)} 

\begin{document}
\maketitle
\section*{Abstract} 
Upon starting  a  collective endeavour, it is important to understand   your partners' preferences and how strongly they commit to a common goal. Establishing a prior commitment or agreement in terms of posterior benefits and consequences from those engaging in it  provides an important mechanism for securing cooperation. Resorting to methods from Evolutionary Game Theory (EGT),  here we analyse  how prior commitments can also be  adopted as a tool for enhancing coordination when its outcomes exhibit an asymmetric payoff structure, in both pairwise and multiparty interactions. Arguably, coordination is  more complex to achieve than cooperation since there might be several desirable collective outcomes in a coordination problem (compared to mutual cooperation, the only desirable collective outcome in cooperation dilemmas). Our analysis, both  analytically and via numerical simulations,  shows that whether prior commitment would be a viable evolutionary mechanism for enhancing  coordination and the overall population social welfare strongly depends on the collective benefit and severity of competition, and more importantly, how asymmetric benefits are resolved in a commitment deal. Moreover, in multiparty interactions, prior commitments  prove to be  crucial  when  a high level of group diversity is required for optimal coordination. The results are robust for different selection intensities.  Overall, our analysis  provides new insights into the complexity and beauty of behavioral evolution  driven by humans' capacity for commitment, as well as for the design of self-organised and distributed multi-agent systems for ensuring coordination among autonomous agents.
\\

\noindent \textbf{Keywords:} Commitment, Evolutionary Game Theory, Coordination, Technology Adoption. 

\newpage

\section{Introduction}
Achieving a collective endeavour  among  individuals with their own personal interest is an important social and economic challenge in various societies \citep{hardin:1968mm,ostrom1990governing,pitt2012axiomatization,barrett2016coordination,key:Sigmund_selfishnes}.    
From coordinating individuals in the workplace to maintaining cooperative and trust-based  relationship among organisations and  nations, it success is often jeopardised by individual self-interest \citep{barrett2007cooperate,perc2017statistical}.
The study of mechanisms that support the evolution of such collective behaviours has been of great interest in many disciplines, ranging from Evolutionary Biology, Economics, Physics and  Computer Science  \citep{key:novak2006,key:Sigmund_selfishnes,tuyls2007evolutionary,West2007,HanBook2013,perc2017statistical,andras2018trusting,kumar2020evolution}.  
Several mechanisms responsible for the emergence and stability of collective behaviours among such individuals, have been proposed, including kin and group selection, direct and indirect reciprocities, spatial networks, reward and punishment \citep{key:novak2006,West2007,perc2017statistical,okada2020review,skyrm1996}.

Recently, establishing prior commitments   has been proposed as an evolutionarily viable strategy inducing cooperative behaviour in the context of  pairwise and multi-player cooperation  dilemmas \citep{Nese2011-chapter,frank88,HanJaamas2016,Han:2014tl,sasaki2015commitment,arvanitis2019agreement,ohtsuki2018evolutionary}; namely, the Prisoner's Dilemma (PD) \citep{han2013good,hasan2013emergence} and the Public Goods Game (PGG) \citep{Han:2014tl,HanJaamas2016,kurzban2001incremental}. It provides an enhancement  to different forms of  punishment against  inappropriate behaviours and  of rewards to stimulate the appropriate ones \citep{chen2014probabilistic,martinez2015apology,martinez2017agreement,sasaki2015commitment,powers2012punishment,szolnoki2012evolutionary,cimpeanu2019exogenous,wang2019exploring}, allowing ones to efficiently avoid free-riders \citep{hanTom2016synergy,han2015synergy} and resolve the antisocial punishment problem \citep{Han2016AAAI}. 
These works have primarily  focused on modelling prior commitments for improving mutual cooperation among self-interested agents. 
In the context of cooperation  dilemma games (i.e.  PD and PGG), mutual cooperation is the only desirable collective outcome to which  all parties are required to commit if an agreement is to be formed. 
  {The same argument  is applied to other pairwise and multi-player social dilemmas such as the Stag-Hunt and Chicken games, since although the nature of the games  is different from the PD and PGG, mutual cooperation is the only desirable outcome to be achieved \citep{key:santos2006,pacheco:2009aa,skyrm2003}. }
In other contexts such as coordination problems, this is not the case anymore since there might be multiple optimal or desirable collective outcomes and players might have distinct, incompatible   preferences regarding which outcome a mutual  agreement should aim to achieve (e.g. due to asymmetric benefits). Such coordination problems are abundant in nature, ranging from collective hunting and foraging to  international climate change actions and multi-sector coordination  \citep{SantosPNAS2011,ostrom1990governing,barrett2016coordination,ohtsuki2018evolutionary,techadoption2019,skyrm1996,santos2016evolutionary}.


Hence,  we explore how arranging a prior agreement or commitment can be used as a mechanism for enhancing coordination and the population social welfare in this type of coordination problems, in both pairwise and multi-player interaction settings. 
Before individuals embark on a joint venture, a pre-agreement makes the motives and intentions of all parties involved more transparent, thereby enabling an easier coordination of personal interests   \citep{Nese2011-chapter,cohenLevesque90,HanBook2013,han2015synergy}. Although our approach is applicable for a wide range of coordination problems (e.g. single market product investments as described above), we will frame our models within the technology investment strategic decision making problem, allowing us to describe the models clearly. 
Namely, we describe technology adoption games capturing the competitive market and decision-making process among firms adopting new technologies \citep{zhu2003strategic,bardhan2004prioritizing}, with a key parameter $\alpha$ representing how competitive the market is (thus describing how important coordination is). 
Similar to previous commitment models, we will perform theoretical  analysis  and numerical simulations resorting to stochastic methods from Evolutionary Game Theory (EGT) \citep{key:Hofbauer1998,key:sigmund2010}.


We will start by modelling a pairwise technology adoption decision making, where two investment firms (or players) competing within a same product market who need to  make  strategic decision on which technology to adopt \citep{zhu2003strategic,chevalier2011strategic}, a  low-benefit (L) or a high-benefit (H) technology. Individually, adopting H would lead to a larger benefit. However, if both firms invest on H they would end up competing with each other leading to a smaller accumulated benefit than if they could coordinate with each other to choose different technologies. However, given the asymmetry in the benefits in such an outcome, clearly no firm would want to commit to the outcome where its option is L, unless some form of compensation from the one selecting H can be ensured.

We then extend and generalize the pairwise model to a multi-player one, capturing the strategic interaction between more than two investment firms. 
 In the multi-player model,  a key parameter $\mu$  is ascribed to the market demand of high technology, i.e. what is the optimal fraction of the firms in a group to adopt H. We analytically examine how players can be coordinated when there is a market demand for a particular technology. We show that differently from the two-player game, the newly defined parameter $\mu$ leads to  a new kind of complexity when trying to achieve group coordination. When there is a high level of diversity in demand (i.e. intermediate values of $\mu$), as can be seen in different technologies adoption contexts  \citep{beede1998patterns,schewe2015diversity}, introducing prior commitment can lead to significant improvement in the levels of coordination and population  social welfare.

The next section discusses related work, which is followed by a description of our models and details of the EGT methods for analysing them. Results of the analysis and a final discussion will then follow.

\section{Related Work} 

The problem of explaining the emergence and stability  of collective behaviours has been actively addressed in different disciplines \citep{key:novak2006,key:Sigmund_selfishnes}.  
Among other mechanisms, such as reciprocity and costly punishment, 
closely related to our present model is the study of cooperative behaviours and pre-commitment in cooperation dilemmas, for both two-player and multiplayer games  \citep{han2013good, HanJaamas2016,sasaki2015commitment,hasan2013emergence,quillien2020evolution}.  
It has shown that to enhance cooperation commitments need to be sufficiently enforced and the cost of setting up the commitments is justified with respect to the benefit derived from the interactions---both by means of theoretical analysis   and of behavioural experiments  \citep{ostrom1990governing,cherry2013enforcing,kurzban2001incremental,chen1994effects,arvanitis2019agreement}. 
Our results show that this same observation is seen for  coordination problems. 
However, arranging  commitments for enhancing coordination is more complex, exhibiting a larger behavioural space, and furthermore, their outcomes strongly depend on new factors only appearing in coordination problems; namely, a successful commitment deal needs to take into account the fact that multiple desirable collective outcomes exist for which  players  have incompatible preferences; and thus how benefits can  be shared through compensations in order to  resolve the issues of asymmetric benefits, is crucially important
\citep{techadoption2019}.

We moved further by expanding our two-player game in the previous work to a multi-player model, the outcome has shown to be more complex as there are more players involved. We yet again investigated how coordination and cooperation can be improved using prior commitment deal when there are multiple players involved and also when there is a particular market demand 
\citep{techadoption2019}.
Our approach in exploring how implementing prior commitment enhances cooperation dilemma has also been investigated by previous researchers in the past 
\citep{chen1994effects}.
A good level of cooperation was seen in a Public Good Game experiment when there was a binding agreement made during the prior communication stage among members of the group. They hypothesized that if members of a group are allowed to make a pledge (a degree of bindings/commitment) before their actual decisions, they will be able to communicate their intentions and it will overall increase cooperation rate in the population. As predicted, their results clearly   {demonstrate} that making a pledge improves cooperation although the degree of commitment required in the pledge deferentially affected the cooperation rate
\citep{chen1994effects,cherry2013enforcing,kurzban2001incremental}.

There have been several other works studying the evolution of coordination, using the so-called Stag Hunt game, see e.g.  \citep{skyrm2003,pacheco:2009aa,key:santos2006,key:Sigmund_selfishnes}. However, to the best of our knowledge there has been no work studying how prior commitments can be modelled and used for enhancing  the outcome of the evolution of coordination. As our results below show, significant enhancement of coordination and population welfare can be achieved via the arrangement of suitable commitment deals. 

Furthermore, it is noteworthy that commitments have been studied extensively in Artificial Intelligence  and  Multi-agent systems literature, see e.g.  \citep{Singh91intentions_commitments,castelfranchi2010,Chopra09m.p.:multiagent,rzadca2015game,Harrenstein:2007,Winikoff:2007}. 
Differently from our work, these studies utilise commitments  for the purpose of regulating  individual and collective behaviours, formalising different  aspects of commitments (such as norms and conventions) in multi-agent systems. 
However, our results and approach  provide  important new insights into the  design of such  systems as these  require commitments to ensure high levels of efficient collaboration and coordination within a group or team  of agents. For example, by providing suitable agreement deals agents can improve the chance that a desirable collective outcome (which is best for the systems as a whole) is reached even when benefits provided by the outcome are different for the parties involved. 
\section{Models and Methods}
\label{sec:models_and_methods}
In the following  {,} we first describe a two-player technology adoption game then extend it with the  option of  arranging prior commitments before playing the game. We then present a multi-player version of the model,  with and without commitments, too. Then, we describe the methods, which are based on Evolutionary Game Theory for finite populations, which will be used to analyse the  resulting models.
\subsection{Two-player Tech Adoption (TD) Game}
\subsubsection{Two-player TD Without Commitments}
\label{subsec:PD}
We consider the scenario that two firms (players) compete for the same product market, and they need to  make a (strategic) decision on which technology to invest on, a  low-benefit (L) or a high-benefit (H) technology. 
  The outcome of the interaction  can be described in terms of costs and benefits of investments by the following payoff matrix (for row player): 

{\small 
\begin{equation}
\label{eq:PMatrix}
 \bordermatrix{~ & H & L\cr
                  H & \alpha b_H - c_H & b_H - c_H \cr
                  L & b_L - c_L & \alpha b_L - c_L  \cr
                 } =  \bordermatrix{~ & H & L\cr
                  H &  a & b \cr
                  L & c &  d  \cr
                 }, 
\end{equation}
}where $c_L$, $c_H$ 
and $b_L$,  $b_H$ ($b_L \leq b_H$) represent the costs and benefits of investing on L and H, respectively; $\alpha \in (0,1)$ indicates the competitive level of the product market: the firms receive a partial benefit if they both choose  to  invest   on  the same technology.  Collectively, the smaller $\alpha$ is (i.e. the higher the market competitiveness), the more important that the firms coordinate to choose different technologies. For simplicity, the entries of the payoff matrix are denoted by $a, \ b, \ c, \ d$, as above.  We have $b > a$ and $c > d$. Without loss of generality, we assume that H would generate a greater net benefit, i.e.  $c = b_L - c_L < b_H - c_H = b$. 

Note that although we describe our model in terms of  technology adoption decision making,  it is  generally applicable to many other coordination problems for instance wherever there are strategic investment decisions to make (in   competitive  markets of any products) \citep{zhu2003strategic,chevalier2011strategic}.

\subsubsection{Two-player TD in Presence of Commitments}
\label{2Comit}
We now extend the model allowing players to have the option to  arrange a prior commitment before a TD interaction. 
A commitment proposal is to ask the co-player to adopt a different technology. That is, a strategist  intending to use H (resp., L) would ask the co-player to adopt L (resp., H). We denote these commitment proposing strategies as HP and LP, respectively. 
Similarly to previous models of commitments (for PD and PGG)  \citep{han2013good,Han:2014tl}, to make  the commitment deal reliable, a proposer  pays an arrangement  cost  $\epsilon$. 
  If the co-player agrees with the deal, then the proposer assumes that the opponent will adopt the agreed choice, yet there is no guarantee that this will actually be the case. Thus whenever a co-player refuses to commit, HP and LP would play H in the game. 
  When the co-player accepts the commitment  though later does not honour it, she has to compensate the honouring co-player at   a personal cost $\delta$.  
  
Differently from previous models on PD and PGG where an agreed outcome  leads to the same payoff for all parties in the agreement (mutual cooperation benefit), in the current model  such an outcome would  lead to different payoffs for those involved. Therefore, as part of the agreement, HP would compensate after the game an amount $\theta_1$ to accepted player that honours the agreement; while LP would request a compensation $\theta_2$ from such an  accepted co-player. 
  
Besides HP and LP, we consider a minimal model with the following (basic) strategies in this commitment version:   
\begin{itemize} 
\item Non-proposing  acceptors, HC and LC, who always commit when being proposed a commitment deal wherein  they are willing to adopt any technology proposed (even when it is different from their intended choice),  honour  the adopted agreement, but do not propose a commitment themselves. They play their intended choice, i.e. H and L, respectively,  when there is no agreement in place;
\item Non-acceptors, HN and LN, who do not accept commitment,  play their intended choice during the game, and do not propose commitments;
  \item Fake committers, HF and LF,  who accept a commitment proposal yet play the choice opposite to what has been agreed  whenever the game takes place. These players assume that they can exploit the commitment proposing  players without suffering the consequences \footnote{Compared to cooperation dilemmas such as PD and PGG, fake strategies make less sense in the context of coordination games since they would not earn the temptation payoff by adopting a different choice from what being agreed. Moreover, in the presence of an agreement, players obtain an additional compensation when adopting the disadvantageous choice (i.e. L). We will keep  the fake strategies in the analysis of pairwise games for confirmation of these intuitions but will exclude them from multi-player settings for simplicity, without  being detrimental to the results.  }.  
\end{itemize} 
Note that similar to the commitment models for the PD game \citep{han2013good}, some possible strategies have been excluded from the analysis since they are   dominated by at least one of the strategies in any configuration of the game: they can be omitted without changing the outcome of the analysis.  For example, those who propose a commitment (i.e. paying a cost $\epsilon$) but then do not honour (thus have to pay the compensation when facing a honouring acceptors)  would be dominated by the corresponding non-proposers. 

Together the model consists of  eight strategies that define the following payoff matrix, capturing the average payoffs that each strategy will receive upon interaction with one of the other seven  strategies (where we denote  $\lambda = \theta_1 + \theta_2$, $\lambda_1 = b-\epsilon-\theta_1$, $\lambda_2 = c-\epsilon+\theta_2$, $\lambda_3 =  a-\epsilon+\delta$ and $\lambda_4 =  d-\epsilon+\delta$, just for the sake of clear representation) 

{\small
\begin{equation} 
 \label{payoff_matrix_share_cost}
\bordermatrix{~ & \text{HP}  & \text{LP}  & \text{HN} & \text{LN} & \text{HC} & \text{LC} & \text{HF} & \text{LF} \cr
  \text{HP}& \frac{b+c-\epsilon}{2} &\frac{2b-\epsilon-\lambda}{2}& a &b & \lambda_1 & \lambda_1 & \lambda_3 & \lambda_3  \cr
\text{LP} &\frac{2c-\epsilon+\lambda}{2}& \frac{b+c-\epsilon}{2}& a &b & \lambda_2 &\lambda_2 & \lambda_4 & \lambda_4  \cr
 \text{HN} &a & a & a & b & a  & b & a & b   \cr
 \text{LN} &c & c & c & d & c  & d & c & d   \cr
 \text{HC} &c+\theta_1 & b-\theta_2 &a  & b  & a & b & a & b  \cr
 \text{LC} &c+\theta_1 & b-\theta_2 &c  & d  & c & d & c & d  \cr
 \text{HF} &a-\delta  & d-\delta  &a  & b  & a & b & a & b  \cr
 \text{LF} &a-\delta  & d-\delta  &c  & d  & c & d & c & d  \cr
}.
\end{equation}
}
Note that when two commitment proposers interact  only one of them will need to pay the cost of setting up the commitment. Yet, as either one of them can take this action they pay this cost only half of the time (on average). In addition, the average payoff of HP when interacting with LP is given by $\frac{1}{2}(b - \epsilon - \theta_1 + b - \theta_2 ) = \frac{1}{2}(2b - \epsilon - \theta_1 - \theta_2 )$. When two HP players interact, each receives $\frac{1}{2}(b - \epsilon - \theta_1 + c + \theta_1 ) = \frac{1}{2}(b+c - \epsilon)$. 

We say that \textit{an agreement is fair} if both parties obtain the same benefit when they honour it  (after having taken into account the cost of setting up the agreement). For that, we can show that $\theta_1$ and $\theta_2$ must satisfy $\theta_1 =\frac{b-c-\epsilon}{2} $ and $\theta_2 =\frac{b-c+\epsilon}{2} $, and thus, both parties obtain $\frac{b+c-\epsilon}{2}$.  Indeed, they can be achieved by comparing the payoffs of HP and HC when they interact, i.e. $b -  {\epsilon} -\theta_1 = c + \theta_1$, where solving this equation we would obtain $\theta_1 =\frac{b-c-\epsilon}{2}$. 

With these conditions it also ensures that the payoffs of HP and LP when interacting with each other are equal. Our analysis below will first focus on whether and when the fair agreements can lead to improvement in terms of coordination and the  overall social welfare (i.e. average population payoff). We will discuss how different kinds of agreements (varying $\theta_1$ and $\theta_2$) affect the outcome, with additional results provided in Appendix. 

\subsection{Multi-Player Tech Adoption (TD) Game}
\subsubsection{Multi-player TD Without Commitments}
We now describe  a $N$-player ($N > 2$) version of the TD model. Again, as before, we will introduce the model in the context of technology investment market decision making.  In a group (of size $N$) with  $k$ players of type $H$ (i.e., $N-k$ players of type $L$), the expected payoffs of playing $H$ and $L$ can be written as follows
\begin{equation}
\label{eq:PayoffAMulti}
\begin{split} 
\Pi_H(k)&= \alpha_{H}(k)b_H-c_H, \\ 
\Pi_L(k)&=\alpha_{L} (k)b_L-c_L,
\end{split}
\end{equation}  
where $\alpha_{H}(k)$ and $\alpha_{L}(k)$ represent the fraction of the benefit obtained by H and L players, respectively, which depend on the composition of the group, $k$.  For two-player TD, both are equal to $\alpha$. To generalize  for $N$-player TD interactions, they should also depend on the  demand for high technology (H) in the group, describing what is the maximal number of players in the group that can adopt H without reducing their benefit due to competition. Let us  denote this number  by $\mu$ (where $1 \leq \mu \leq N$). For example, intermediate values of $\mu$ indicate a high level of group diversity is needed for optimal coordination. When $\mu = N$, it means there is a significant market demand of the high benefit technology so that all firms can adopt it without leading to competition. 

Hence, we  define 
\begin{equation}
  \alpha_{H}(k)=\begin{cases}
    1, & \text{if $k\le \mu$},\\
    \frac{ \alpha_{1} \mu}{k} & \text{otherwise},
  \end{cases}
\end{equation}

\begin{equation}
  \alpha_{L}(k)=\begin{cases}
    1, & \text{if $ k \geq \mu$},\\
    \frac{\alpha_{2} (N-\mu)}{N-k} & \text{otherwise}.
  \end{cases}
\end{equation}
The rationale of these definitions is that whenever $k \leq \mu$, full benefits from adopting H can be obtained, and moreover, if $k > \mu$, the larger $k$ the stronger the competition is among H adopters. Similarly for L adopters. 
The parameters $\alpha_{1}$ and $\alpha_{2}$ stand for the  intensities of competition for investing in H and in L, respectively. For simplicity we assume in this paper $\alpha_{1}=\alpha_{2}=\alpha$. 
Note that for $N=2$ we recover the two-player model  given in Equation \eqref{eq:PMatrix}  {, given  that the current $\alpha$ is scaled (by 2) compared to the value of   $\alpha$ in the pairwise game, solely for the purpose of a clear presentation}.

The optimal group payoff is achieved when there are exactly $\mu$ players adopting H and the rest adopting L, leading to an average payoff for each member given by 
$$
A:=\frac{\mu(b_{H}-c_{H})+(N-\mu)(b_{L}-c_{L})}{N}.
$$
\subsubsection{Multi-player TD in Presence of Commitments}
We can define the $N$-player game version with prior commitments in a similar fashion as in the two-player game. Commitment proposing strategists (i.e. HP and LP players) will propose before an interaction that the group will play the optimal arrangement (so that every player obtains an average payoff $A$). For simplicity, we  assume that the committed players adopt the fair agreement, i.e. every member will obtain the same payoff after compensation is made to those adopting L. As such, we don't need to consider who will adopt H or L, as all would receive the same payoff at the end. 
Moreover, whenever a player in the group refuses to commit, commitment proposers will adopt H. 
Details of payoff calculation will be provided in Results section (cf. Table 1). 


\subsection{ Evolutionary Dynamics} 
\label{subsec:evolution} 
In this work,  we will perform theoretical  analysis and numerical  simulations   (see next section)  using   EGT methods for finite populations \citep{key:novaknature2004,key:imhof2005,key:Hauert2007}.  Let $Z$ be the size of the population. In such a setting,  individuals'  payoff represents their \emph{fitness} or social \emph{success}, and  evolutionary dynamics is shaped  by social learning \citep{key:Hofbauer1998,key:Sigmund_selfishnes}, whereby the  most successful individuals  will tend to be imitated more often by the other individuals. In the current work, social learning is modelled using  the so-called pairwise comparison rule \citep{traulsen2006}, a  standard approach in EGT,  assuming  that an individual  $A$ with fitness $f_A$ adopts the strategy of another individual  $B$ with fitness $f_B$ with probability $p$ given by the Fermi function, 
$$p_{A, B}=\left(1 + e^{-\beta(f_B-f_A)}\right)^{-1}.$$
The parameter  $\beta$ represents  the `imitation strength' or `intensity of selection', i.e., how strongly the individuals  base their decision to imitate on fitness difference between themselves and the opponents. For $\beta=0$,  we obtain the limit of neutral drift -- the imitation decision is random. For large $\beta$, imitation becomes increasingly deterministic.
 
In the absence of mutations or exploration, the end states of evolution are inevitably monomorphic: once such a state is reached, it cannot be escaped through imitation. We thus further assume that, with a certain mutation probability,  an individual switches randomly to a different strategy without imitating  another individual.  In the limit of small mutation rates, the dynamics will proceed with, at most, two strategies in the population, such that the behavioural dynamics can be conveniently described by a Markov Chain, where each state represents a monomorphic population, whereas the transition probabilities are given by the fixation probability of a single mutant \citep{key:imhof2005,key:novaknature2004,key:Hauert2007}. The resulting Markov Chain has a stationary distribution, which characterises the average time the population spends in each of these monomorphic end states.   {It has been shown to have a range of applicability which goes well beyond the strict limit of very small mutation (or exploration) rates  \citep{key:Hauert2007,key:Sigmund_selfishnes,key:Hanetall_AAMAS2012,key:sigmund2010,randUltimatum}.} 

Before describing how to calculate this stationary distribution, we need to show how payoffs are calculated, which differ for two-player and N-player settings, as below.
\begin{itemize}
\item[$\bullet$] \textbf{Average Payoff for the Two Player Game }\\
Let $\pi_{ij}$ represent the payoff obtained by strategist $i$ in each pairwise interaction with strategist $j$, as defined in the payoff matrices in Equations \eqref{eq:PMatrix} and \eqref{payoff_matrix_share_cost}. Suppose there are at most two strategies in the population, say, $  {x}$ individuals using $i$ ($0\leq   {x} \leq Z$) and ($Z-  {x}$)  individuals using $j$. Thus the average payoff of the individual that uses $i$ or $j$ can be written respectively as follows 
\begin{equation} 
\label{eq:PayoffA}
\begin{split} 
\Pi_i(  {x}) &=\frac{(  {x}-1)\pi_{ii} + (Z-  {x})\pi_{i,j}}{Z-1},\\
\Pi_j(  {x}) &=\frac{  {x}\pi_{j,i} + (Z-  {x}-1)\pi_{j,j}}{Z-1}.
\end{split}
\end{equation} 
\item[$\bullet$] \textbf{Expected Payoff in The Multiplayer Game} \\
In  the case of $N$-player interactions, suppose the population includes $x$ individuals of type $i$ and $Z-x$ individuals of type $j$. The probability to select $k$ individuals of type $i$ and $N-k$ individuals of type $j$, in $N$ trails, is given by the hypergeometric distribution as follows \citep{key:Sigmund_selfishnes,GokhalePNAS2010} 
$$
H(k,N,x,  {Z})=\frac{\binom{x}{k}\binom{Z-x}{N-k}}{\binom{Z}{N}}
$$
Hence, in a population of $x$ $i$-strategists and $(Z-x)$ $j$ strategists, the average payoff of $i$ and $j$ are given by 
\begin{equation}
\begin{split} 
\Pi_{ij}(  {x})&=\sum_{k=0}^{N-1}H(k,N-1,x-1,Z-1)\pi_{ij}(k+1) = \sum_{k=0}^{N-1}\frac{\binom{x-1}{k}\binom{Z-x}{N-1-k}}{\binom{Z-1}{N-1}} \pi_{ij}(k+1), \\
\Pi_{ji}(  {x})&=\sum_{k=0}^{N-1}H(k,N-1,x-,Z-1)\pi_{ji}(k)=\sum_{k=0}^{N-1}\frac{\binom{x}{k}\binom{Z-1-x}{N-1-k}}{\binom{Z-1}{N-1}} \pi_{  {ji}}(k).
\end{split}
\end{equation}
\end{itemize}
Now, for both two-player and $N$-player settings,  the probability to change the number $  {x}$ of individuals using strategy A by $\pm$ one in each time step can be written as \citep{traulsen2006} 
\begin{equation}
\label{eq:Traone} 
T^{\pm}(k) = \frac{Z-  {x}}{Z} \frac{  {x}}{Z} \left[1 + e^{\mp\beta[\Pi_i(  {x}) - \Pi_j(  {x})]}\right]^{-1}.
\end{equation}
The fixation probability of a single mutant with a strategy $i$ in a population of $(  {Z}-1)$ individuals using $j$ is given by \citep{traulsen2006,key:novaknature2004}
\begin{equation} 
\label{eq:fixprob} 
\rho_{j,i} = \left(1 + \sum_{i = 1}^{  {Z}-1} \prod_{j = 1}^i \frac{T^-(j)}{T^+(j)}\right)^{-1}.
\end{equation} 
Considering a set  $\{1,...,q\}$ of different strategies, these fixation probabilities determine a transition matrix $M = \{T_{ij}\}_{i,j = 1}^q$, with $T_{ij, j \neq i} = \rho_{ji}/(q-1)$ and  $T_{ii} = 1 - \sum^{q}_{j = 1, j \neq i} T_{ij}$, of a Markov Chain. The normalised eigenvector associated with the eigenvalue 1 of the transposed of $M$ provides the stationary distribution described above \citep{key:imhof2005}, describing the relative time the population spends adopting each of the strategies. 

\paragraph{Risk-dominance} 
An important measure to determine the evolutionary dynamic of a given strategy is its risk-dominance against others. For the two strategies $i$ and $j$, risk-dominance is a criterion which determine which selection direction is more probable: an $i$ mutant is able to fixating in a homogeneous population of agents using $j$ or a $j$ mutant fixating in a homogeneous population of individuals playing $i$. In the case, for instance, the first was more probable than the latter then we say that $i$ is \emph{risk-dominant} against $j$ \citep{key:novaknature2004,key:Sigmund_selfishnes} which holds for any intensity of selection and in the limit for large population size $Z$ when 
\begin{equation}
\label{eq:compare_fixprob_cond_largeN}
\sum_{k=1}^{N}\Pi_{i,j}(k)\geq \sum_{k=0}^{N-1} \Pi_{j,i}(k) 
\end{equation}
This condition is applicable for both two-player games, $N = 2$,  and when N-player games with $N > 2$ \citep{key:Sigmund_selfishnes,GokhalePNAS2010}.   {It will  allow us to derive analytical conditions such as when commitment proposing is an evolutionarily viable strategy, being risk-dominant against all other strategies in the population. }

\begin{table}[!t]
    \begin{tabular}{|p{11cm}|c|}
    \hline
    \textbf{Parameters description}  & \textbf{Notation} \\
     \hline \hline
        Cost of investing in high technology, H & $c_H$ \\
      \hline
    Cost of investing in low technology, L & $c_L$\\
    \hline
      Benefit  of investing in high technology, H  & $b_H$\\
   \hline
    Benefit  of investing in low technology, L & $b_L$\\
     \hline
    Competitive level of the market & $\alpha$\\
       \hline
   Group size (in N-player TD games) & $N$\\
       \hline
    Optimal number of H-adopters in a group of N players& $\mu$\\
       \hline
     Cost of arranging a commitment & $\epsilon$\\
       \hline
   Compensation paid by dishonouring commitment acceptors & $\delta$ \\
         \hline
     Compensation paid by HP to honouring commitment acceptors & $\theta_1$\\
       \hline
     Compensation paid to LP by commitment acceptors & $\theta_2$\\
  \hline
 \end{tabular}
 \caption{  {List of parameters in the models.}}
\label{tbl:BModel}
\end{table}

\section{Results}
We will first describe results for two-player games, then proceeding to provide those for the $N$-player version.   {Table \ref{tbl:BModel} summarizes the key parameters in both versions, for ease of following. }
\subsection{Two-player TD Game Results } 
\subsubsection{Analytical Conditions for the Viability of Commitment Proposers} 
To begin with, using  the conditions given in  Equation \ref{eq:compare_fixprob_cond_largeN}, we obtain that if $$\theta_1 + \theta_2 < b-c$$ then HP is  risk-dominant (see Methods) against  LP. Otherwise, LP is risk-dominant  against HP. 

Similarly, we  derive the conditions regarding the commitment parameters for which HP and LP are evolutionarily viable strategies, i.e. when they are risk-dominant against all other non-proposing ones. Indeed,  HP and LP are risk-dominant against all other six non-proposing strategies, respectively, if and only if

\begin{equation}
\begin{split}
\epsilon &< \min\{ b+c-2a, 3b-c-2d, \frac{3b-c-2a-4\theta_1}{3},  
\frac{3b-c-2d-4\theta_1}{3}, \frac{b+c-2a+4\delta}{3},  \frac{b+c-2d+4\delta}{3}\}, \\
\epsilon &< \min\{ b+c-2a, 3b-c-2d, \frac{3c- b - 2a +4\theta_2}{3},  \frac{3c - b- 2d+4\theta_2}{3}, \frac{b+c-2a+4\delta}{3},  \frac{b+c-2d+4\delta}{3}\}.
\end{split}
\end{equation} 
Note that each element in the $\textit{min}$ expressions above corresponds to the condition for one of the six non-proposing strategies HN, LN, HC, LC, HF, LF, respectively. 

Thus, we can derive the conditions for $\theta_1$, $\theta_2$ and $\delta$: 
\begin{equation}
\begin{split}
\theta_1 &< \frac{1}{4}\left(3b-c - 3\epsilon -  2\max\{a,d\}\right), \\
\theta_2 &> \frac{1}{4}\left(b-3c+ 3\epsilon +  2\max\{a,d\}\right), \\
\delta &> \frac{1}{4}\left(3\epsilon-b-c + 2\max\{a,d\}\right).
\end{split} 
\end{equation}
In particular, for fair agreements, i.e.  $\theta_1 = (b-c-\epsilon)/2$ and $\theta_2 = (b-c+\epsilon)/2$, we obtain
\begin{equation}
\label{eq:conditions_risk_dom}
\begin{split}
&\epsilon <   b + c - 2\max\{a,d\}, \\
\delta &> \frac{1}{4}\left(3\epsilon-b-c + 2\max\{a,d\}\right).
\end{split}
\end{equation} 
It is  because $3b-c-2d > b + c - 2\max\{a,d\}$, which is due to $b > c$ and $\max\{a,d\} \geq d$.

\begin{figure*}
\centering
\includegraphics[width=\linewidth]{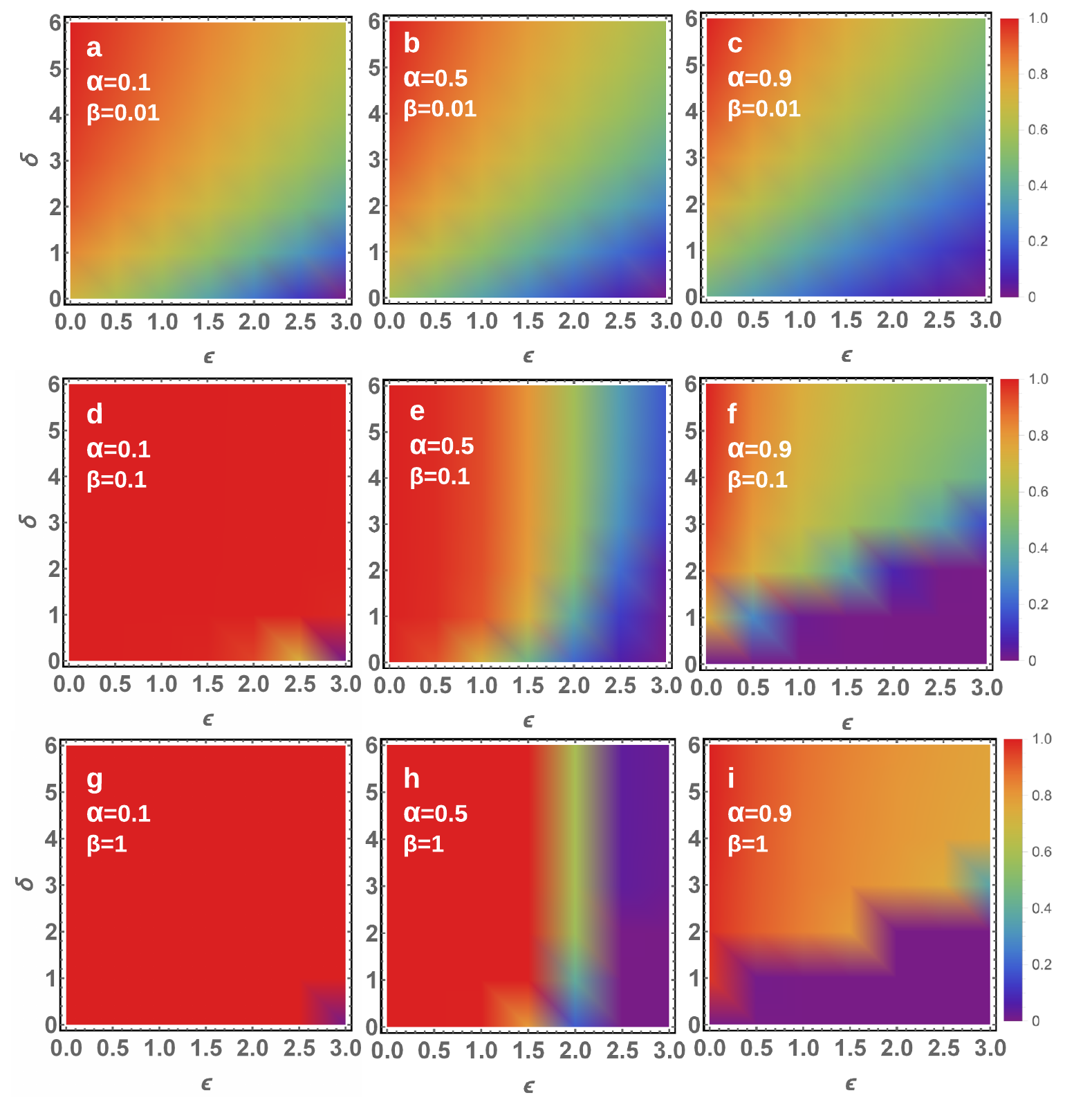} 

\caption{\textbf{Total frequency of commitment strategies (i.e. sum of the frequencies of HP and LP), as a function of $\epsilon$ and $\delta$}, for different values of $\alpha$ and $\beta$. Primarily, the commitment proposing strategies dominate the population whenever $\epsilon$ is sufficiently small and $\delta$ is sufficiently large. Furthermore, the smaller $\alpha$,  these commitment strategies dominate for a wider range of $\epsilon$ and $\delta$, especially when $\alpha$ is smaller.   {These observations are robust for different values of  $\beta$. Nevertheless,  a larger $\beta$  leads to a greater frequency of commitment proposing strategies where they are evolutionarily viable, and a lower frequency otherwise. }   Parameters: in all panels  $c_H = 1$, $c_L = 1$, $b_L = 2$ (i.e. $c = 1$), and  $b_H = 6$ (i.e. $b = 5$).  Other parameters:  $\beta = 0.01$ in the first, $\beta = 0.1$ in the second  and $\beta = 1$ in the third row; population size $Z = 100$; Fair agreements are used, where $\theta_1$ and $\theta_2$ are given by $\theta_1 = (b-c-\epsilon)/2$ and $\theta_2 = (b-c+\epsilon)/2$. }
\label{fig:Fig_vary_eps_delta}
\end{figure*}

In general,  these conditions indicate that for commitments to be a viable option for improving coordination, the cost of arrangement $\epsilon$ must be sufficiently small while the compensation associated with the contract needs to be sufficiently large (see already Figure \ref{fig:Fig_vary_eps_delta} for numerical validation).  
Furthermore, for the first  condition to hold, it is necessary that $b + c > 2\max\{a,d\}$. It means that the total payoff of two players when playing the TD game is always greater when they can coordinate to choose different technologies, than when they both choose the same technology. 
 
 Moreover, the conditions in Equation \ref{eq:conditions_risk_dom} can be expressed in terms of $\alpha$ and the costs and benefits of investment, as follows (see again the payoff matrices in Equation \ref{eq:PMatrix})
\begin{equation*}
\begin{split}
&\alpha <     {\frac{1}{2}} + \min\{ \frac{c_H+b_L - c_L - \epsilon}{2 b_H},  \frac{c_L+b_H - c_H - \epsilon}{2 b_L}  \}, \\
&\alpha <     {\frac{1}{2}}  + \min\{ \frac{c_H+b_L - c_L - 3\epsilon + 4\delta}{2b_H},  \frac{c_L+b_H - c_H -  3\epsilon + 4\delta}{2b_L}  \},
\end{split}
\end{equation*} 
which can be rewritten as 

\begin{equation}
\label{eq:conditions_risk_dom_alpha}
\begin{split}
&\alpha <     {\frac{1}{2}}  + \min\{ \frac{c_H+b_L - c_L - \max\{\epsilon,3\epsilon -4\delta\}}{2b_H},  \frac{c_L+b_H - c_H - \max\{\epsilon,3\epsilon -4\delta\}}{2b_L}  \}.
\end{split}
\end{equation} 
 This condition indicates under what condition of the market competitiveness and the costs and benefits of investing in available technologies, commitments can be an evolutionarily viable mechanism. 
 Intuitively, for given costs and benefits of investment (i.e. fixing $c_L$, $c_H$, $b_L$, $b_H$), a larger cost of arranging a (reliable) agreement, $\epsilon$, leads to a smaller threshold of $\alpha$ where commitment is viable. Moreover, given a commitment system (i.e. fixing $\epsilon$ and $\delta$), assuming similar costs of investment for the two technologies, then a larger  ratio of the benefits obtained from the two technologies, $b_H/b_L$, leads to a smaller upper bound for $\alpha$ for which commitment is viable.  

\begin{figure*}
\centering
\includegraphics[width=\linewidth]{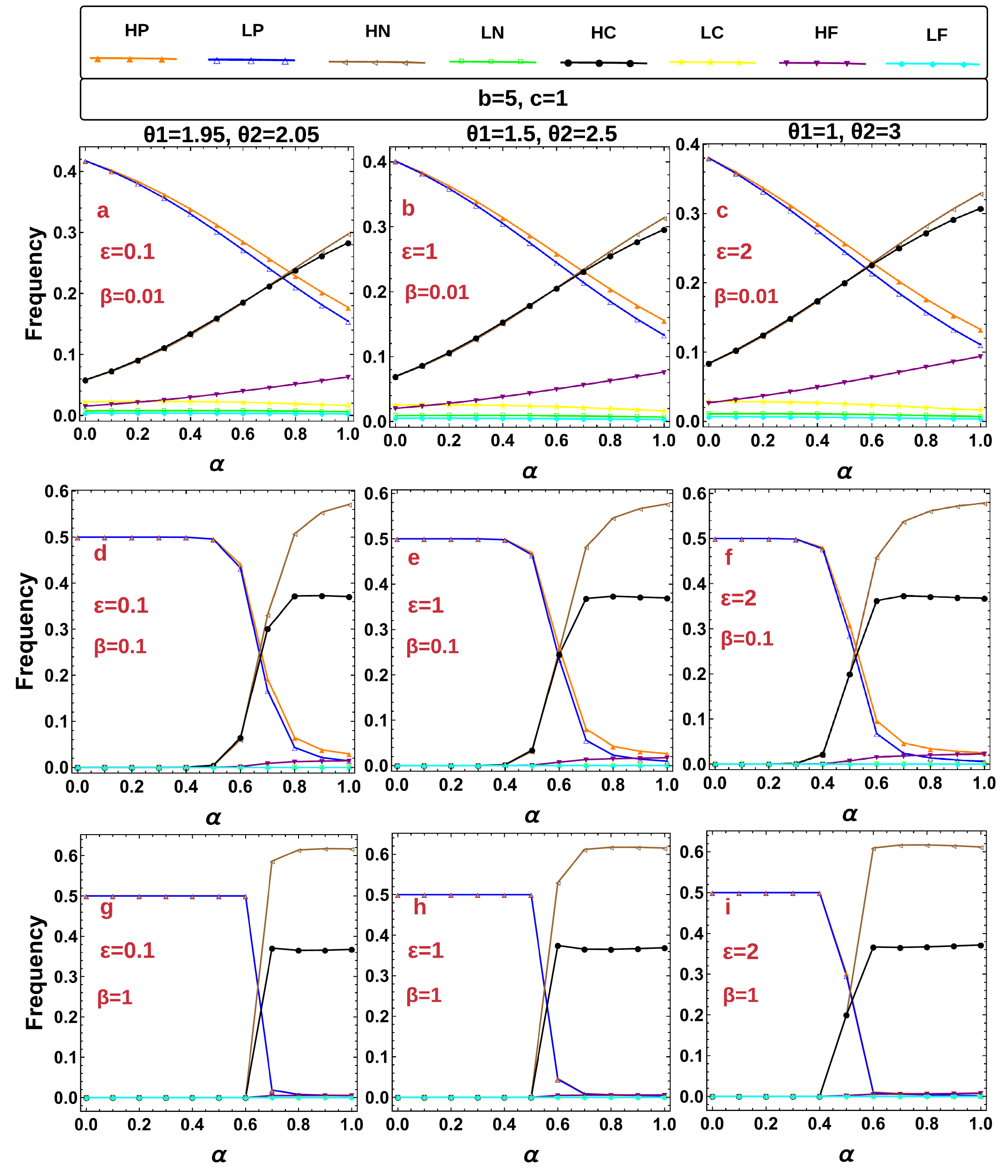} 
\caption{\textbf{Frequency of the eight strategies, HP, LP, HN, LN, HC, LC, HF and LF, as a function of $\alpha$}, for different values of $\epsilon$ and $\beta$.  In general, the commitment proposing strategies HP and LP dominate the population when $\alpha$ is small while HN and HC dominate when $\alpha$ is sufficiently large in all cases, which is robust for different values of intensity of selection, $\beta$. The HN and HC dominate the population as the market competition decreases (i.e. when  $\alpha$ increases). Larger values of $\beta$ increase the difference between strategies' frequencies but do not change the outcomes in general. Parameters: in all panels  $c_H = 1$, $c_L = 1$, $b_L = 2$ (i.e. $c = 1$), $b_H = 6$ (i.e. $b = 5$). Other parameters:  $\delta = 6$; $\beta = 0.01, 0.1$  and 1; population size $Z = 100$; Fair agreements are used, where $\theta_1$ and $\theta_2$ are given by $\theta_1 = (b-c-\epsilon)/2$ and $\theta_2 = (b-c+\epsilon)/2$. }
\label{fig:Fig_vary_alpha}
\end{figure*}

\begin{figure}
\centering
\includegraphics[width=\linewidth]{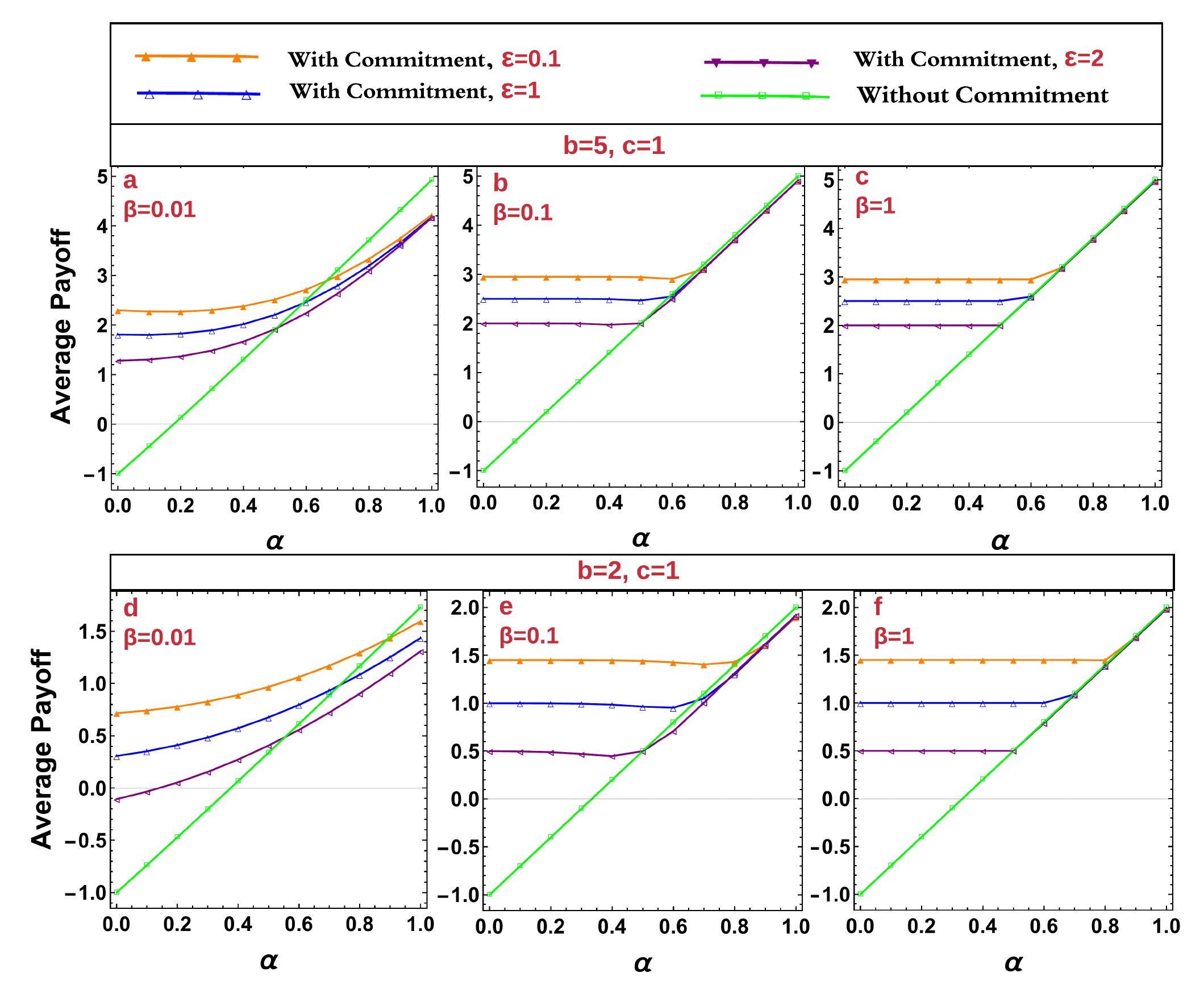} 
\caption{\textbf{Average population  payoff as a function of $\alpha$, when commitment is absent  and  when it is present, for different values of $\epsilon$ and $\beta$}. We observed that when $\alpha$ is small, significant improvement in terms of the average population payoff can be achieved through prior commitment. When $\alpha$ is sufficiently large, commitment leads to   {no} improvement or might even be detrimental for social welfare, especially when $\beta$ is small. That is, at $\alpha = 0.7$ in panel a and $\alpha = 0.9$ in panel d, without commitment will be more beneficial. Parameters: in all panels $c_H = 1$, $c_L = 1$, $b_L = 2$ (i.e. $c = 1$); in panel a, b and c) $b_H = 6$ (i.e. $b = 5$) with $\beta = 0.01, \ 0.1$ and $1$ respectively. Also, in panels d, e and f) $b_H = 3$ (i.e. $b = 2$) with $\beta = 0.01, \ 0.1$  and $1$ respectively; Other parameters:  $\delta = 6$; population size $Z = 100$; Fair agreements are used, where $\theta_1$ and $\theta_2$ are given by $\theta_1 = (b-c-\epsilon)/2$ and $\theta_2 = (b-c+\epsilon)/2$.   }
\label{fig:compare_with_vs_without_commitment}
\end{figure}

 Remarkably, our numerical analysis  below (see already Figure \ref{fig:Fig_vary_alpha}) shows that the condition in Equation \ref{eq:conditions_risk_dom_alpha} accurately predicts the threshold of $\alpha$ where commitment proposing strategies (i.e. HP and LP) are highly abundant  in the population, leading to improvement in terms of the average population payoff compared to when commitment is absent (Figure \ref{fig:compare_with_vs_without_commitment}).     {For example, when $\epsilon = 0.1, \ 1$ and $2$, the upper bounds for $\alpha$ are $0.658$, $0.583$ and $0.5$, respectively. }

 On the other hand, when $\alpha$ is sufficiently large, little improvement can be achieved, especially when $b_H/b_L$ is large (which is in accordance with the analytical results above).


\subsubsection{Numerical Results for  Pairwise TD game} 
We calculate the stationary distribution in a population of eight strategies, HP, LP, HN, LN, HC, LC, HF and LF, using methods described above. In Figure \ref{fig:Fig_vary_alpha}, we show the frequency of these strategies  as a function of $\alpha$, for different values of $\epsilon$ and game configurations.  In general, the commitment proposing strategies HP and LP dominate the population when $\alpha$ is small while HN and HC dominate when $\alpha$ is sufficiently large even with different values of   {$\beta$} utilized in the comparison. That is, commitment proposing strategies are viable and successful whenever the market competitiveness is high, leading to the need of efficient coordination among the competing players/firms to ensure high benefits. Notably, we observe that the thresholds of $\alpha$ below which HP and LP are dominant, closely  corroborate  the analytical condition described in Equation \ref{eq:conditions_risk_dom_alpha}, in all cases. This observation is also robust for different values of intensity of selection, $\beta$.


This observation is robust for varying commitment parameters, i.e. the cost of arranging commitment, $\epsilon$,  and the compensation cost associated with commitment, $\delta$, see Figure  \ref{fig:Fig_vary_eps_delta}. Namely, we show the total frequency of commitment strategies (i.e. sum of the frequencies of HP and LP) for varying these parameters and for different values of $\alpha$. It can be seen  that, in general, the commitment strategies   dominate the population  whenever $\epsilon$ is sufficiently small and $\delta$ is sufficiently large. This observation is in accordance with previous commitment modelling works for the cooperation dilemma games \citep{han2013good,Han:2014tl,HanJaamas2016}. 
In addition, we  observe that in the current coordination problem, that the smaller $\alpha$ is,  these commitment strategies dominate the population for wider range of $\epsilon$ and $\delta$. Our additional results show that these observations are robust with respect to other game configurations,   {including $\beta$ (comparing the three rows in Figure \ref{fig:Fig_vary_eps_delta}. } 

Now, in order to determine  whether and when commitments can actually lead to meaningful improvement, in Figure \ref{fig:compare_with_vs_without_commitment}, we compare the average population payoff or social welfare when a commitment is present and when it is absent. 
In general, it can be seen that when $\alpha$ is sufficiently small (below a threshold), the smaller it is, the greater improvement of social welfare  is achieved through the presence of a commitment deal. Moreover, the smaller the cost of arranging commitments, $\epsilon$,  the greater improvement is obtained. 
When $\alpha$ is sufficiently large, commitment leads to   {no} improvement or might even be detrimental for social welfare, 
 especially when $b_H/b_L$ is large (which is in accordance with the analytical results above).  The detriment is further increased when $\beta$ is small.
We can observe that the thresholds for which a notable improvement can be achieved is the same as the one for the viability of HP and LP (i.e. as described in Equation \ref{eq:conditions_risk_dom_alpha}). 



\subsection{Multiplayer Game Results}
\subsubsection{Payoff Derivation in $N$-player TD game}
As mentioned above, compared to cooperation dilemmas such as PD and PGG, fake strategies make less sense in the context of coordination games since they would not earn the temptation payoff by adopting a different choice from what being agreed. To focus on the group effect and the effect of the  newly introduced parameter $\mu$, we will consider a population consisting of HP, LP, HN, LN, HC and LC (i.e. excluding fake strategies). As shown in the two-player game analysis, the fake strategies (i.e. HF and LF) are not viable options in TD games and can be ignored. It is equivalent to consider to the full set of strategies with a sufficiently large $\delta$.  

First of all, we derive the payoffs received by each strategy when encountering specific other strategies (see a summary in Table \ref{tab:Tabbl}). Namely, $\Pi_{ij}(k)$ and $\Pi_{ji}(k)$ denote  the payoffs of a strategist of type $i$ and $j$, respectively, in a group consisting of $k$ player of type $i$ and $N-k$ players of type $j$. The first column of the table lists all possible strategies which can be used by player $i$ (focal player), where as the second column shows strategies of co-players (opponents). The third column shows the payoffs of focal players. 

\begin{center}
\begin{table}
\label{tab:Tabbl}
\begin{tabular}{|c|c|c|}
\hline 
\rule[-1ex]{0pt}{3ex} Focal Player ($i$) & Opponent ($j$) &$\Pi_{i,j}(k)$ \\ [1.5ex]
\hline 
\rule[-1ex]{0pt}{3ex} HP, LP & HP, LP& $A - \epsilon/N$ \\   [1.5ex]
\hline 
\rule[-1ex]{0pt}{3ex} HP, LP & HC, LC & $A - \epsilon/k$ \\   [1.5ex]
\hline 
\rule[0.1ex]{0pt}{3ex} HP, LP & HN & $\Pi_{H}(N)$ (for $k < N$)\\  [1.5ex]
\hline 
\rule[0.1ex]{0pt}{3ex} HP, LP & LN & $\Pi_{H}(k)$ (for $k < N$)\\  [1.5ex]
\hline
\rule[-1ex]{0pt}{2.5ex} HN &   {HP, LP, HN, HC}& $\Pi_{H}(N)$ \\ [1.5ex] 
\hline 
\rule[-1ex]{0pt}{2.5ex} HN &   {LN, LC} & $\Pi_{H}(k)$ \\ [1.5ex] 
\hline
\rule[-1ex]{0pt}{2.5ex} LN &   {HP, HN, HC} & $\Pi_{L}(k)$ \\ [1.5ex]
\hline 
\rule[-1ex]{0pt}{2.5ex} LN &   {LN, LC}  & $\Pi_{L}(N)$ \\ [1.5ex]
\hline 
\rule[-1ex]{0pt}{2.5ex} LN & LP  & $\Pi_{L}(k)$ \\ [1.5ex]
\hline 
\rule[-1ex]{0pt}{2.5ex} HC, LC &   {HP, LP} & A (for $k < N$)  \\ [1.5ex]
\hline 
\rule[-1ex]{0pt}{2.5ex} HC &   {HN, HC} &$\Pi_{H}(N)$  \\[1.5ex] 
\hline 
\rule[-1ex]{0pt}{2.5ex} HC &   {LN, LC} &$\Pi_{H}(k)$  \\[1.5ex]
\hline 
\rule[-1ex]{0pt}{2.5ex} LC &   {HN, HC} & $\Pi_{L}(k)$   \\[1.5ex] 
\hline
\rule[-1ex]{0pt}{2.5ex} LC &   {LN, LC} & $\Pi_{L}(N)$   \\[1.5ex] 
\hline 
\end{tabular}
\caption{\label{tab:table-name} Average payoffs of focal strategy $i$ when facing strategy $j$, in a group of $k$ former and $N-k$ latter strategists.}
\label{tab:Tabbl} 
\end{table}
\end{center}
\subsubsection{Analytical conditions for the viability of commitment proposers in N-player TD game}
We now derive the conditions under which  HP is risk-dominant against the rest of strategies. Since we assume fair agreements, the conditions for LP would be  equivalent to those for HP in terms of risk-dominance. 
  {For ease of following the derivations below, we recall that $A$ denotes the optimal group payoff achieved when there are exactly $\mu$ players adopting H and the rest adopting L, that is, 
$A:=\frac{1}{N}\left(\mu(b_{H}-c_{H})+(N-\mu)(b_{L}-c_{L}) \right)$.
}

  {HP} is risk-dominant against HC if
$$
\sum_{k=1}^{N}\Pi_{HP,HC}(k)\geq \sum_{k=0}^{N-1} \Pi_{HC,HP}(k),
$$
which can be written as $$
\sum_{k=1}^{N}\left(A -  \frac{\epsilon}{k}\right)  \geq \Pi_{H}(N)+\sum_{k=1}^{N-1} A.
$$
Hence we obtain 
\begin{equation} 
\epsilon \leq \frac{A-\Pi_{H}(N)}{H_{N}},
\label{eq:riskdomHC}
\end{equation} 
where $H_{N}=\sum_{k=1}^{N} \frac{1}{k}$.
 \begin{figure}
\centering
\includegraphics[width=0.8\linewidth]{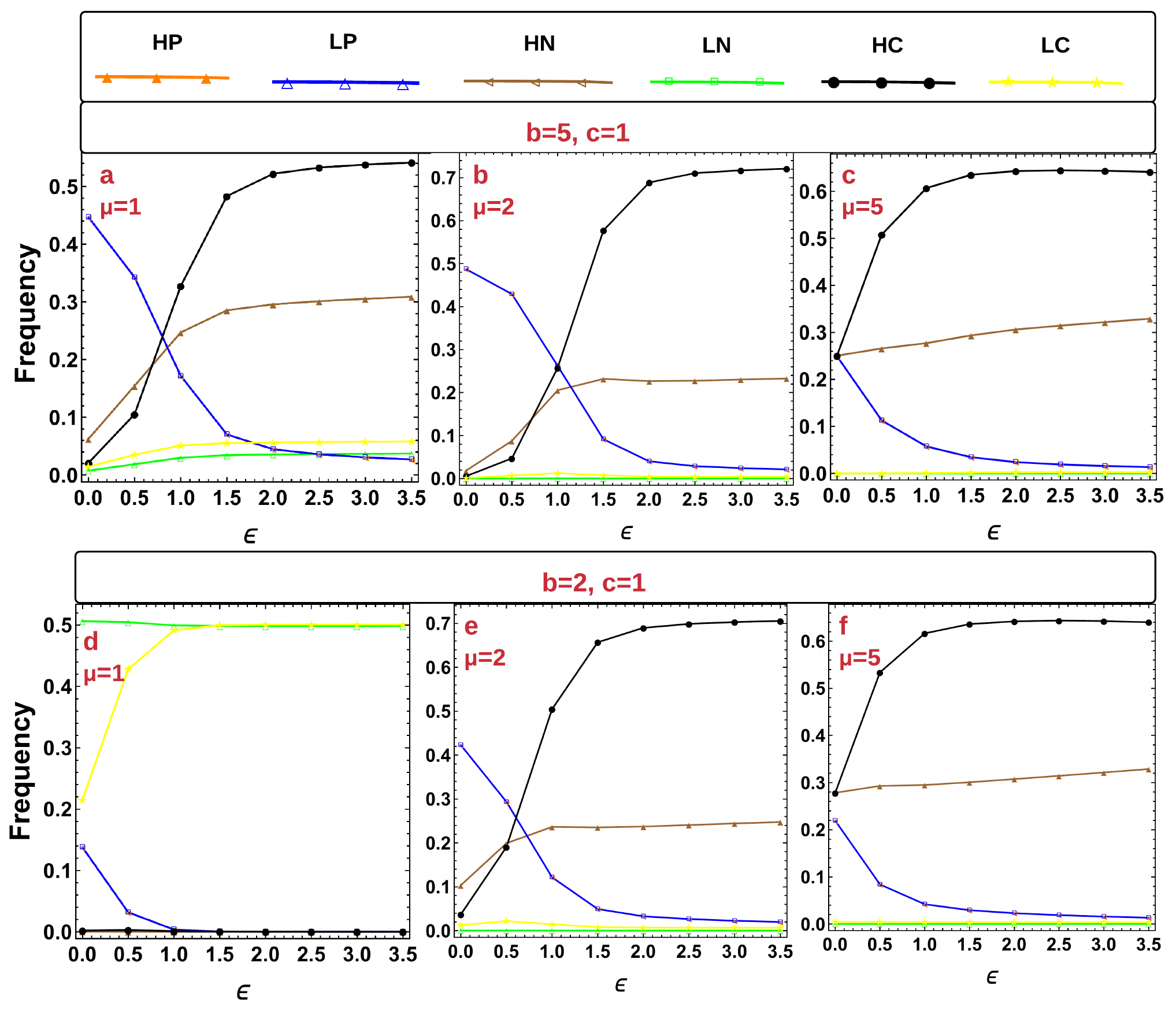}
\caption
{\textbf{Frequency of the six strategies HP, LP, HN, LN, HC and LC, as a function of $\epsilon$ in a N-player game with commitment, for different values of $\mu$}. In the N-player game,  the new parameter $\mu$ describes the market demand for a high technology, which was set to 1 in the pairwise game.
HP and LP have a high frequency for sufficiently small  $\epsilon$ for  $\mu = 2$ in both games, and also when  $\mu = 1$ for the first, easy coordinate situation (first row). When $\mu = 5$, i.e. when all players can adopt H without benefit reduction, HC always dominate and commitment strategies are not successful.  
This means that when there is a need for a diversity of technology adoption, initiating prior commitments to enhance  coordination is  important. Parameters: in panel a, b and c) $b_H = 6$ (i.e. $b = 5$) with $\mu = 1, 2, 5$ respectively. Also, in panel d,e and f) $b_H = 3$ (i.e. $b = 2$) with $\mu=1, 2, 5$ respectively; Other parameters: $N = 5$, $\beta =  0.1$; $\alpha = 0.5$; $c_H = 1$, $c_L = 1$, $b_L = 2$ (i.e. $c = 1$); .} 
\label{fig:Fig_varying_mu_multiplyer}
\end{figure}

Similarly,  HP is risk-dominant against LC if 
\begin{equation} 
\label{eq:riskdomLC}
\epsilon \leq \frac{A-\Pi_{L}(  {0})}{H_{N}}.
\end{equation} 
For risk-dominance of HP against HN, 
$$
\sum_{k=1}^{N}\Pi_{HP,HN}(k)\geq \sum_{k=0}^{N-1} \Pi_{HN,HP}(k),
$$
which equivalently can be written as
$$
A - \frac{\epsilon}{N} \geq \Pi_{H}(N),
$$
or, 
\begin{equation} 
\label{eq:riskdomHN}
\epsilon \leq N \big(  A-\Pi_{H}(N)\big).
\end{equation}
%
 Finally, HP is risk-dominant against LN if 
$$
\sum_{k=1}^{N}\Pi_{HP,LN}(k)\geq \sum_{k=0}^{N-1} \Pi_{LN,HP}(k),
$$
which can be rewritten  as
$$
A - \frac{\epsilon}{N} + \sum_{k=1}^{N-1}\Pi_{H}(k)\geq \sum_{k=0}^{N-1} \Pi_{L}(k),
$$
  {
or
\begin{equation}
\label{eq:riskdomLN}
\epsilon \leq N \left( A + \sum_{k=1}^{N-1}\Pi_{H}(k) - \sum_{k=0}^{N-1} \Pi_{L}(k) \right).
 \end{equation}
}
%
%
In short, in order for commitment proposers to be risk-dominant against all other strategies, it requires that $\epsilon$ is sufficiently small, namely, smaller than minimum of the right hand sides of Equations (\ref{eq:riskdomHC})-(\ref{eq:riskdomLN}).    

\begin{figure}
\centering
\includegraphics[width=0.8\linewidth]{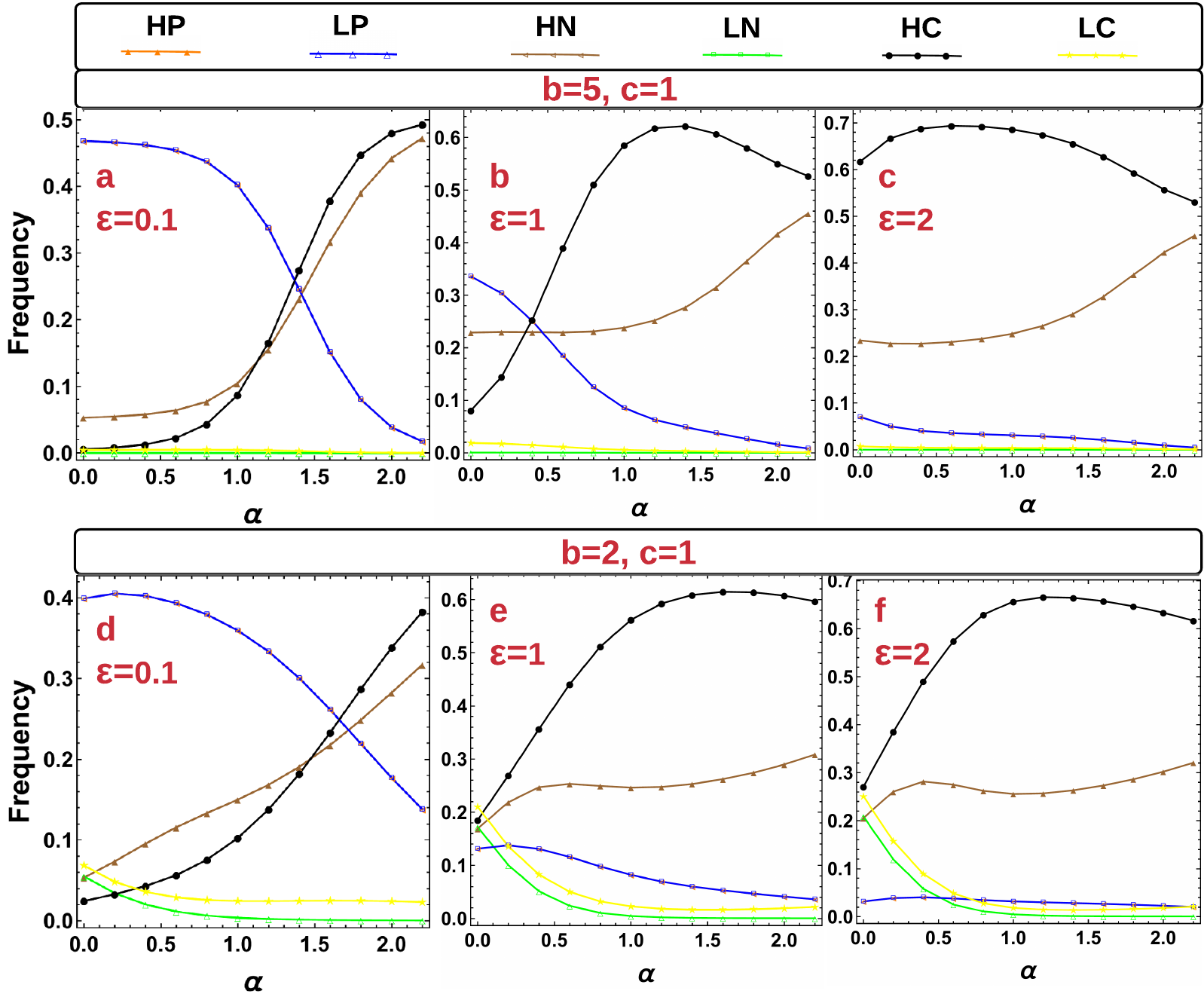} 
\caption{\textbf{Frequency of the six strategies HP, LP, HN, LN, HC and LC, as a function of $\alpha$ in a multiplayer game with commitment}, for different values of $\epsilon$ and also two different game configurations.
In general, the commitment proposing strategies (HP and LP) decrease in frequency for increasing $\alpha$. They dominate over other strategies for sufficiently small $\alpha$ and $\epsilon$.  
That is, it is more beneficial to engage in a prior commitment deal when the market competition is fierce and the cost of arranging the commitment is very minimal. Parameters: in all panels $c_H = 1$, $c_L = 1$, $b_L = 2$ (i.e. $c = 1$); in panel a, b and c) $b_H = 6$ (i.e. $b = 5$) with $\epsilon = 0.1, \ 1$ and  2, respectively. Also, in panel d, e and f: $b_H = 3$ (i.e. $b = 2$) with $\epsilon = 0.1, 1$ and $2$ respectively; Other parameters: $N = 5$, $\beta =  0.1$; $\mu =2$.}
\label{fig:Fig_varying_alpha_multiplyer}
\end{figure}

\begin{figure}
\centering
\includegraphics[width=0.9\linewidth]{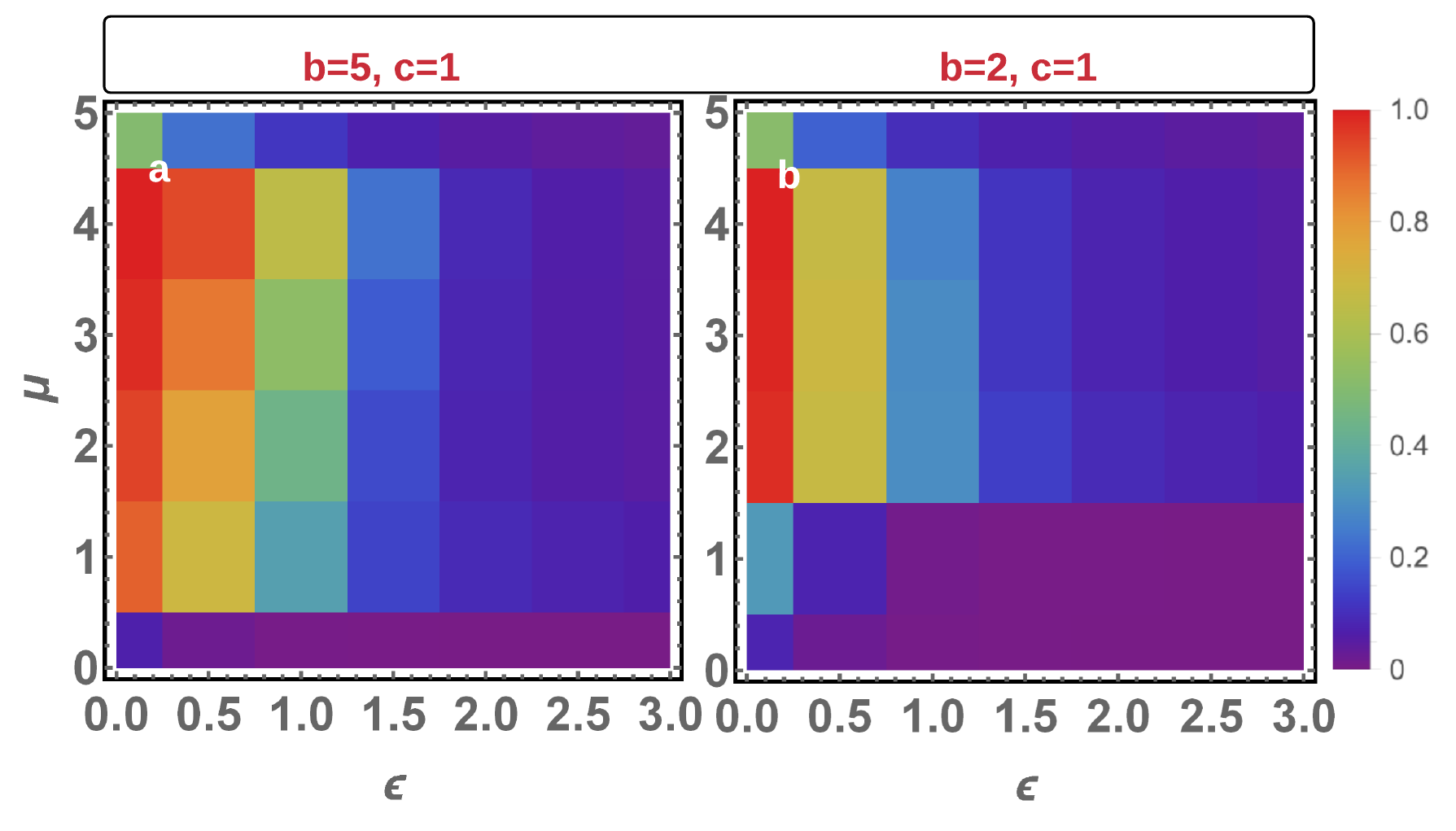}
\caption
{\textbf{Total frequency of commitment proposing strategies HP and LP as a function of $\mu$ and $\epsilon$}. In general, the commitment proposing strategies are most successful for intermediate values of $\mu$, especially for a sufficiently small  cost of arranging prior commitment $\epsilon$.  Parameters: in  all panels, $c_H = 1$, $c_L = 1$ (i.e. $c = 1$), $b_L = 2$. In panel a), $b_H = 6$ (i.e. $b = 5$) and in panel b) $b_H =  3$ (i.e. $b = 2$). Other parameters: $N = 5$, $\beta =  0.1$; $\alpha =0.5$.}
\label{fig:Fig_varying_freq_hplp}
\end{figure}


\begin{figure}
\centering
\includegraphics[width=\linewidth]{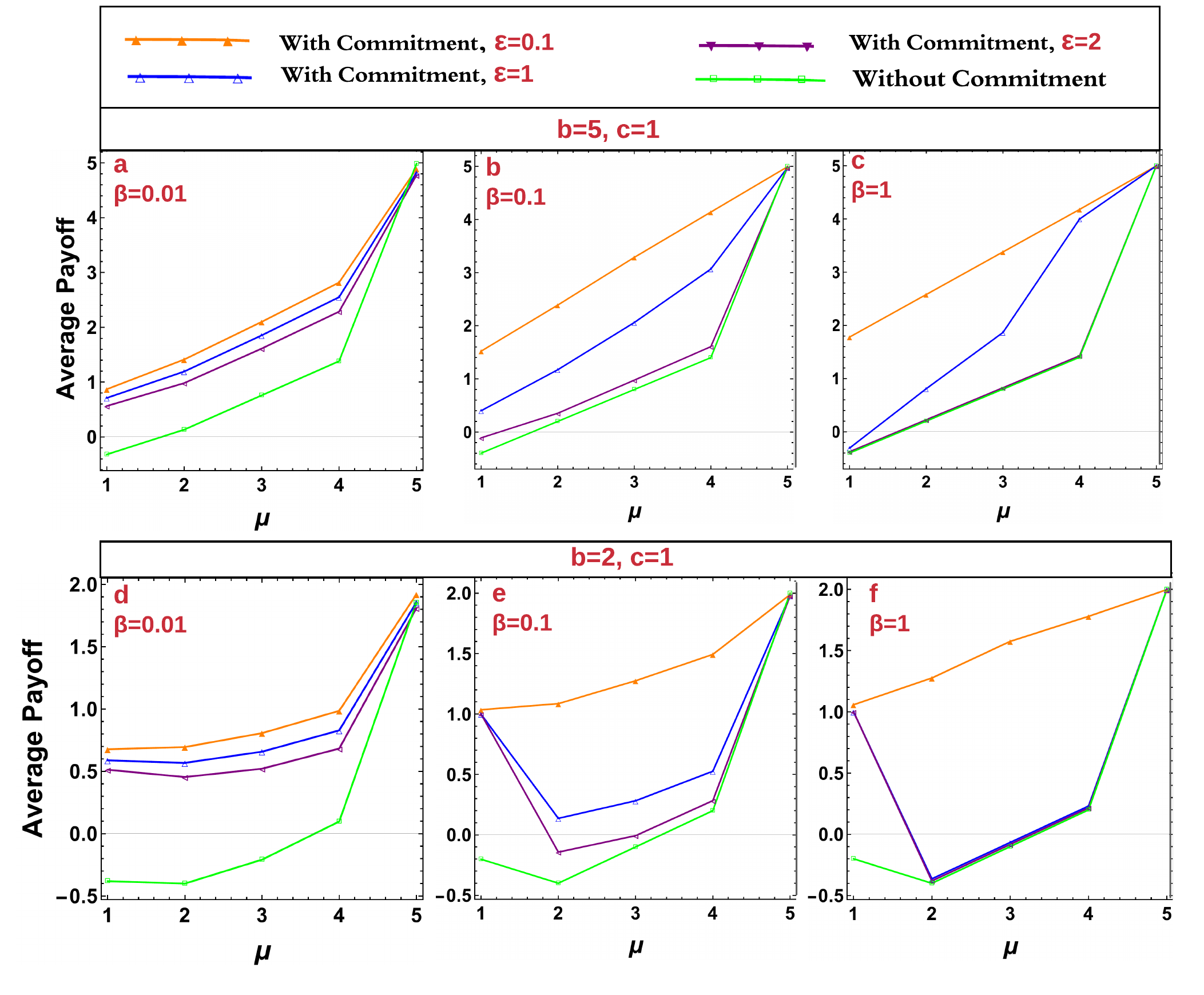}
\caption
{\textbf{Average population payoff (social welfare) as a function of $\mu$ with different values of $\epsilon$, showing when commitment is absent against  when it is present.} We compare results for different values of $\beta$ in two game configurations. We observe that whenever $\mu < 5$ (i.e. when there is a need for coordination to avoid competition in the group), arranging a prior commitment is beneficial to the population social welfare. Parameters:  in panel a, b and c) $b_H = 6$ (i.e. $b = 5$), in panel d,e and f) $b_H = 3$ (i.e. $b = 2$). Other parameters: $N = 5$, $\alpha = 0.5$,  $c_H = 1$, $c_L = 1$, $b_L = 2$}
\label{fig:Commit_Noncommit_DifferentBeta}
\end{figure}

 \subsection{Numerical Results for N-player TD game}
 We  compute stationary distributions in a population of six strategies HP, LP, HN, LN, HC and LC, for the N-player TD game, using the payoffs in Table 1 and the  Methods described above. 
To begin with, in Figure  \ref{fig:Fig_varying_mu_multiplyer} (see also Figure \ref{fig:riskdomNplayer-NumConfirmation} in Appendix), we provide numerical validation for the analytical  conditions obtained in the previous section regarding when commitment proposing strategies are evolutionarily  viable strategies (being risk-dominant against others). Similar to  the pairwise TD game, we observe that there is a threshold for $\epsilon$ below which it is the case. 
Moreover,  Figure \ref{fig:Fig_varying_alpha_multiplyer} shows that the frequencies of  these strategies (HP and LP) decrease for increasing $\alpha$. They dominate the population whenever  $\epsilon$ is sufficiently small (e.g. $\epsilon = 0.1$ and $1$).  
That is, it is more beneficial to engage in a prior commitment deal when the market competition is harsher  (i.e. small $\alpha$).  
These results are  robust for different intensities of selection (see Figure  \ref{fig:Alpha_With_DifferentBeta} in Appendix). 
In general, our results confirm the similar observations regarding the effects of $\epsilon$ and $\alpha$ on the evolutionary outcomes obtained in the  pairwise game above. 

We now focus on understanding the effect of the new parameter in the N-player game, $\mu$, on the evolutionary outcomes. Recall that $\mu$ indicates  the  demand for high technology (H) in the group, describing what is the maximal number of players in the group that can adopt H without reducing their benefit due to competition.  Figure \ref{fig:Fig_varying_mu_multiplyer} shows the effect of different values of $\mu$ on the frequency or evolutionary success of all strategies as a function of $\epsilon$. When $\mu$  is small to intermediate, and the cost of arranging prior commitment is also small, the commitment proposing strategies are dominant. This suggests that arranging prior  commitments might be more  beneficial in such  instances. These results also imply that  $\mu$  is very essential in determining when commitment should be initiated. Apparently, the greater need for a group mixture or market diversity of technologies,  indicating a more difficult coordination situation, the greater need for the utilization of commitment to enhance coordination  among players is.
This observation is even more evident in Figure \ref{fig:Fig_varying_freq_hplp}, where we examine the success of commitment for varying $\mu$ and  $\epsilon$, in regards to  two different game configurations. It can be observed that an intermediate value of $\mu$ leads to the highest frequency of commitment strategies, especially in the more difficult coordination situation (i.e. the right panel). 

We now closely examine  the gain in terms of social welfare improvement when using prior commitments.  As shown in Figure \ref{fig:Commit_Noncommit_DifferentBeta}, whenever $\mu < N $ ($N = 5$), i.e. there is a need to coordinate among the group players to avoid competition that induces benefit reduction, prior commitments lead to increase of social welfare. This increase is more significant in the more difficult coordination situation  (i.e. the lower row) and when the cost of arranging commitment is low, which is also slightly more significant for intermediate values of $\mu$ and higher values of intensity of selection, $\beta$. 


\section{  {Conclusions and Further Discussion}} 
We have  described in this paper novel evolutionary game theory models showing  how prior commitments can be adopted as an efficient mechanism for enhancing coordination, in both pairwise and multi-player interactions. For that, we described technology adoption  (TD) games where technology investment firms would achieve the best collective outcome if they can coordinate with each other to adopt a mixture of different technologies. To this end, a parameter $\alpha$ was used to capture the competitiveness level of the product market and how beneficial it is to achieve coordination, while another parameter $\mu$ to capture the optimal coordination mixture or diversity of technology adopters in a group (in the pairwise case, we assume the optimal mixture is where two firms adopt different technologies to avoid conflict). 

In the coordination settings, there are multiple desirable  outcomes and players have distinct preferences in terms of which outcome should be agreed upon, thus  leading to a larger behavioural space than in the context of cooperation dilemmas \citep{han2013good,HanJaamas2016,Han:2014tl,sasaki2015commitment,hasan2013emergence}.
We have shown that whether commitment is  a viable mechanism for promoting the evolution of   coordination, strongly depends on $\alpha$: when $\alpha$ is sufficiently small, prior commitment is highly abundant leading to significant improvement in terms of  social welfare (i.e. population avarage  payoff), compared to when commitment is absent.
Importantly, we have derived the analytical condition for the threshold of $\alpha$ below which the success of commitments is guaranteed, for both pairwise and multi-player TD games. Furthermore, moving from pairwise to a multi-player setting, it was shown that $\mu$ plays an important role for the success of commitment strategies as well. In general, when $\mu$ is intermediate, equivalent to a high level of diversity in group choices, arranging prior commitments proved to be highly important. It led to significant improvement in terms of social welfare, especially  in a harsher coordination situation.

In the main text, we have considered that a fair agreement is arranged. In the Appendix (Figure \ref{fig:Fig_vary_theta1_theta2}), we have shown that whenever  commitment proposers are allowed to freely choose which deal to propose to their co-players, our results show that, in a highly competitive market (i.e. small $\alpha$), commitment proposers should be strict (i.e. sharing less benefits),  while when the market is less competitive, commitment proposers should be more generous.

  {In both pairwise and multi-player coordination settings, our analysis has shown that the cost of arranging agreement must be sufficiently small, to be justified for the cost and benefit of coordination. This is in line with previous works in the context of PD and PGG \citep{han2013good,HanJaamas2016,Han:2014tl}. It is due to the fact that those who refuse to commit can escape sanction or compensation. Solutions to this problem have been proposed in the context of PD and PGG, namely, to combine commitment with peer punishment, intention recognition, apology or social exclusion to address non-committers   \citep{hanTom2016synergy,han2015synergy,Han:2014tl,martinez2017agreement,quillien2020evolution} or to delegate the costly process of arranging commitment to an external party \citep{cherry2013enforcing,cherry2017refundable}. Our future work will investigate how to combine prior commitments with such mechanisms to provide  a more adaptive and efficient  approach for  coordination enhancement in complex systems.
  }

  {Prior commitments and agreements have been used extensively in the context of  distributed and self-organizing multi-agent systems, for modelling and engineering a desirable correct behaviour, such as cooperation, coordination and fairness  \citep{Singh91intentions_commitments,Chopra09m.p.:multiagent,Winikoff:2007}. These works however do not consider the dynamical  aspects of the systems  nor under what conditions for instance regarding the relation between  costs and benefits of coordination and those of arranging a reliable commitment, commitment proposing strategies can actually promote a high level of desirable system  behaviour. Thus, our results  provide important insights into the design of such distributed and self-organizing (adaptive) systems to ensure high levels of coordination, in both pairwise and multi-party interactions \citep{bonabeau,pitt2012axiomatization}.  
}

In future work,  we will consider how commitments can solve more complex collective problems, e.g. in a technological innovation race \citep{han2019modelling},  bargaining games \citep{zisisSciRep2015,randUltimatum}, climate change actions \citep{barrett2007cooperate,santos2020picky} and cross-sector coordination \citep{santos2016evolutionary}, where there might be a large number of desirable outcomes or  equilibriums, especially when the number of players in an interaction increases \citep{duong2015expected,GokhalePNAS2010}. 

  {Overall,} our work  has demonstrated that commitment is a viable tool for promoting the evolution of diverse collective behaviours among self-interested individuals, beyond the  context of cooperation dilemmas where there is only one desirable collective outcome \citep{skyrm1996,barrett2007cooperate}.    {It thus  provides new insights into the complexity and beauty of behavioral evolution  driven by humans' capacity for commitment \citep{frank88,Nese2011-chapter}. }

\section{Acknowledgements}

 T.A.H. is  supported by a Leverhulme Research Fellowship (RF-2020-603/9). T.A.H and A.E. are also supported by  Future of Life Institute (grant RFP2-154).

\newpage
\section{Appendix} 
\subsection{Results for different values of  $\theta_1$ and $\theta_2$}
\begin{figure*}
\centering
\includegraphics[width=\linewidth]{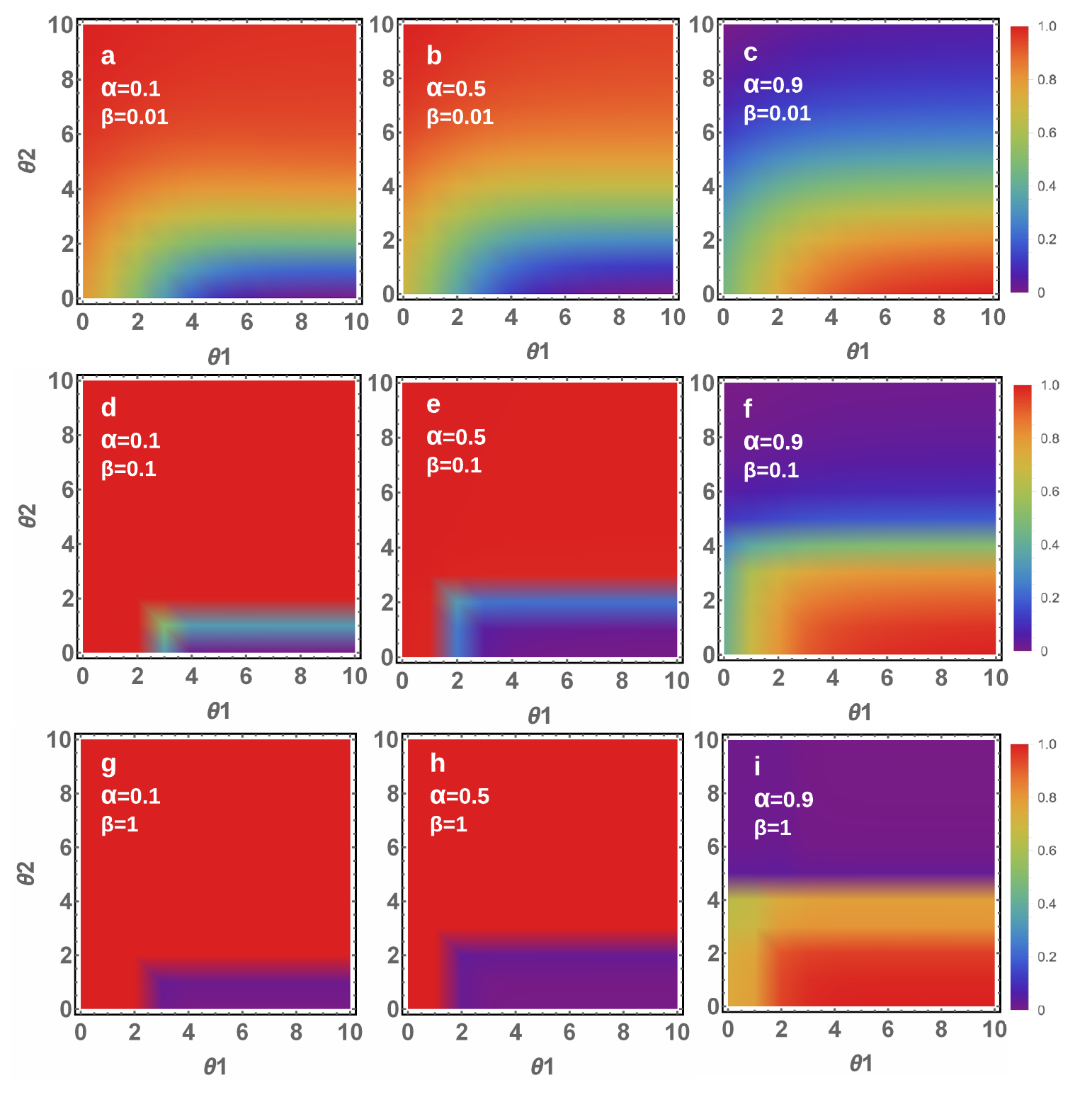} 
\caption{\textbf{Average population payoff as a function of  $\theta_1$ and $\theta_2$}, for different values of $\alpha$ and $\beta$   {(for pairwise TD games)}. When $\alpha$ is small (panels a and b), the highest average payoff is achieved when $\theta_1$ is sufficiently small and $\theta_2$ is sufficiently large, while for large $\alpha$ (panel c), it is the case  when $\theta_1$ is sufficiently large and $\theta_2$ is sufficiently small. Figure 4 also shows that for a small value of $\beta$, the highest average payoff is achieved when $\alpha$ is very minimal compared to other panels with higher value of $\beta$ (compare panel a, d and g). Parameters: in all panels  $c_H = 1$, $c_L = 1$, $b_L = 2$ (i.e. $c = 1$), and  $b_H = 6$ (i.e. $b = 5$). Other parameters:  $\delta = 4, \ \epsilon = 1$; $\beta = 0.01, \ 0.1$ and $1$; population size $Z = 100$. }
\label{fig:Fig_vary_theta1_theta2}
\end{figure*}

 In the main text, we assume that a fair agreement is always arranged. We  consider here what would happen if  HP and LP can personalise  the commitment deal they want  to propose, i.e.  any $\theta_1$ and $\theta_2$ can be proposed  (instead of always being fair). Namely, Figure \ref{fig:Fig_vary_theta1_theta2} shows the average population  payoff varying these  parameters, for different values of $\alpha$. We observe that when $\alpha$ is small, the highest average payoff is achieved when $\theta_1$ is sufficiently small and $\theta_2$ is sufficiently large, while for large $\alpha$, it is reverse for the two parameters. That is, in a highly competitive market (i.e. small $\alpha$), commitment proposers should be strict (HP keeps sufficient benefit while LP requests sufficient payment, from their commitment partners), while when the market is less competitive (i.e. large $\alpha$), commitment proposers should be more generous (HP proposes to give a larger benefit while LP requests a smaller payment, from their commitment partners).  Our results confirm that this observation is robust for different values of $\epsilon$, $\delta$ and $\beta$.

\subsection{Numerical confirmation of risk-dominant conditions in the N-player game}
\begin{figure}
\includegraphics[width=\linewidth]{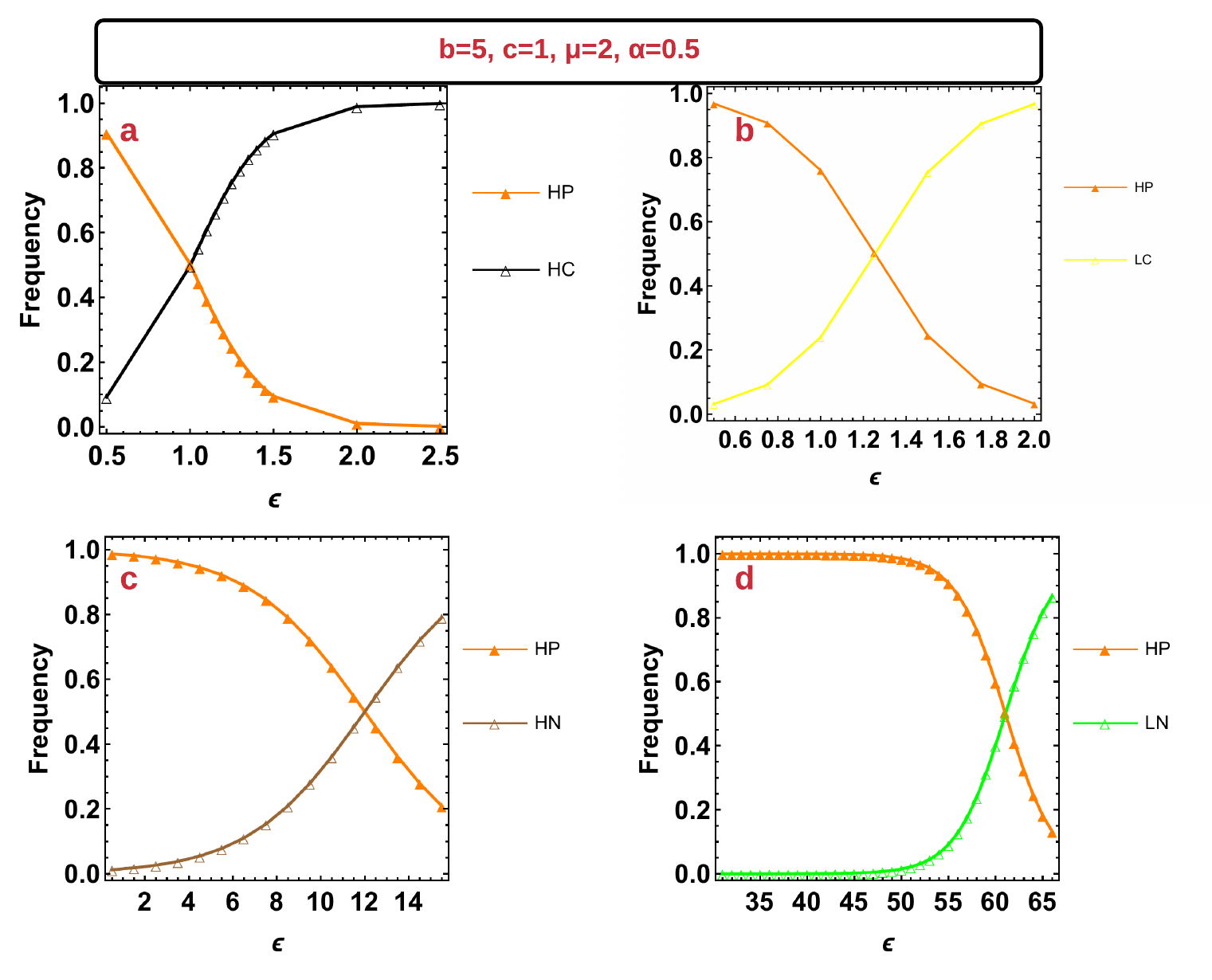}
\caption
{\textbf{Validation for  the analytical  conditions under which HP is risk dominant against strategy HC, LC, HN and LN, see main text.} In all cases, with a small value of $\epsilon$, the HP strategy dominated other players. This result of this figure is in   {close} accordance with our equations derived above.   {Namely, the risk-dominance thresholds of $\epsilon$ for HP (LP) playing against HC, LC, HN and LN, are, 1.05, 1.31, 12.0 and 58.75, respectively. We notice a small difference between numerical and theoretical  results, since the latter ones are approximated for larger population sizes.   } Parameters: in all panels,   {$N = 5$},  $c_H = 1$, $c_L = 1$, $b_H = 6$ (i.e. $b = 5$), $\mu = 2$,  and $\alpha = 0.5$.}
\label{fig:riskdomNplayer-NumConfirmation}
\end{figure}

See Figure \ref{fig:riskdomNplayer-NumConfirmation} for numerical results confirming the risk-dominant conditions in the N-player game in the main text. 

\begin{figure}[H]
\centering
\includegraphics[width=\linewidth]{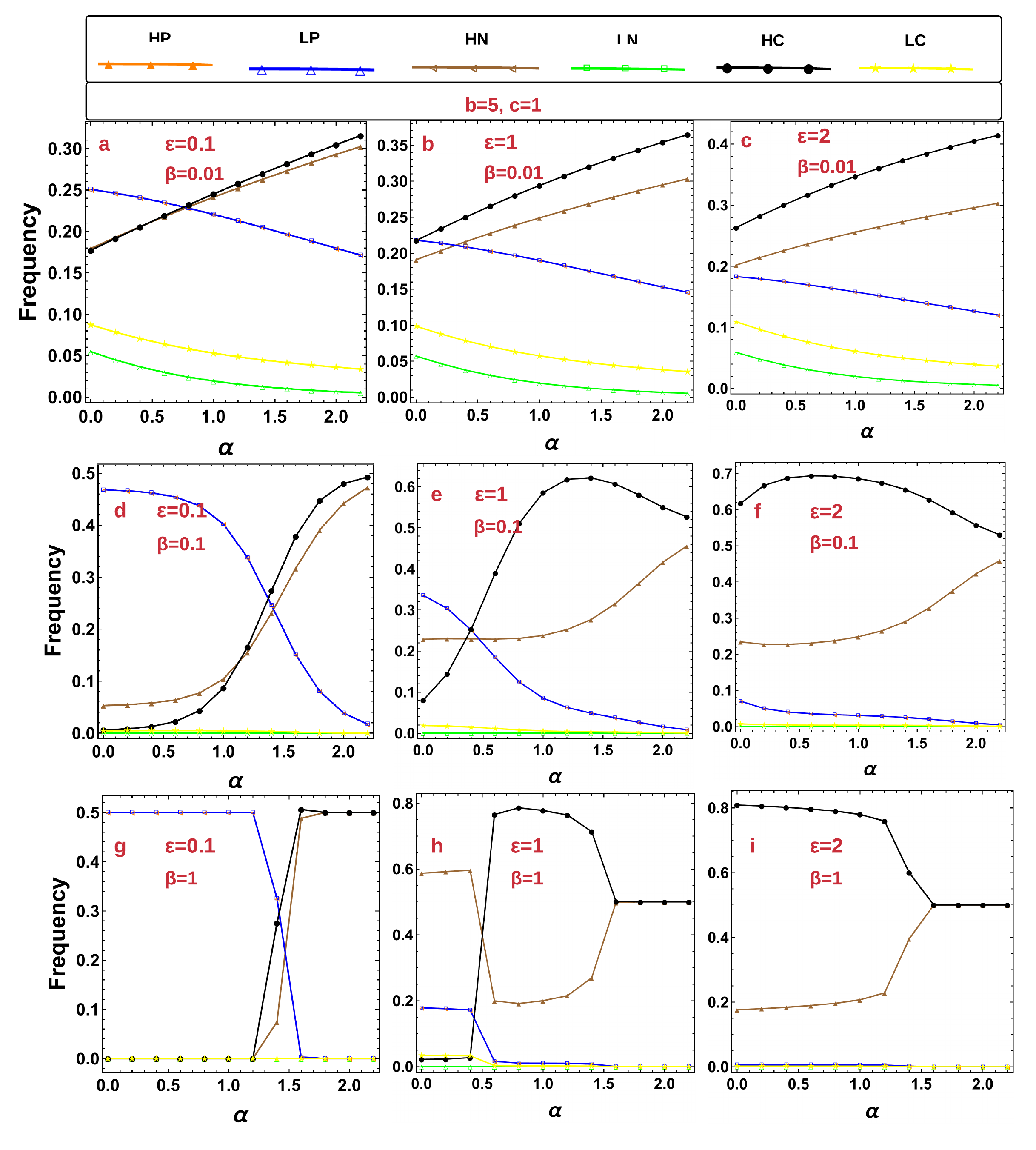}
\caption
{\textbf{Frequency of six strategies HP, LP, HN, LN, HC and LC,  as a function of $\alpha$ and for different values of $\epsilon$ and $\beta$.} The commitment proposing strategies HP and LP dominate the population when the values of $\alpha$ and $\epsilon$ are sufficiently small, in all cases of $\beta$. Furthermore, as the value of $\epsilon$ increases, the non-proposing strategies dominate the population. Parameters: in all panels $c_H = 1$, $c_L = 1$, $b_L = 2$ (i.e. $c = 1$), $b_H = 6$ (i.e. $b = 5$); Other parameters:   {$N = 5$}, $\epsilon = 0.1, \ 1, \ 2$; $\beta = 0.01, \ 0.1, \ 1$.}
\label{fig:Alpha_With_DifferentBeta}
 \end{figure}

  {
\subsection{Results for other  intensities  of selection in the N-player game}
 Figure \ref{fig:Alpha_With_DifferentBeta} confirms similar observations for other values of intensity of selection ($\beta$) in the N-player TD game, as compared to Figure \ref{fig:Fig_varying_alpha_multiplyer} in the main text. 
}

\newpage

\bibliographystyle{apalike}
\bibliography{biblio} 

\end{document}